\documentclass[conference]{IEEEtran}
\IEEEoverridecommandlockouts
\usepackage{cite}
\usepackage{amsmath,amssymb,amsfonts}
\usepackage{algorithmic}
\usepackage{graphicx}
\usepackage{textcomp}
\usepackage{xcolor}

\usepackage{multirow}
\usepackage{enumitem, array}
\usepackage{caption}
\usepackage{subcaption}
\usepackage[export]{adjustbox}
\usepackage{hyperref}
\usepackage{gensymb}

\graphicspath{{images/}}

\def\loss{{\mathcal L}}

\def\C{{\bf C}}
\def\c{{\bf c}}
\def\f{{\bf f}}
\def\X{{\bf X}}
\def\x{{\bf x}}
\def\y{{\bf y}}
\def\Y{{\bf Y}}
\def\T{{\bf T}}
\def\t{{\bf t}}

\def\bphi{{\boldsymbol \phi}}
\def\btheta{{\boldsymbol \theta}}

\def\D{{\mathcal{D}}}

\def\Dcc{{\mathcal{D}_{\mathrm{c} \mathrm{\hat c}}}}
\def\Dx{{\mathcal{D}_{\mathrm{x}}}}
\def\Dxx{{\mathcal{D}_{\mathrm{x} \mathrm{\hat x}}}}
\def\Dtt{{\mathcal{D}_{\mathrm{t} \mathrm{\hat t}}}}
\def\Dt{{\mathcal{D}_{\mathrm{t}}}}

\newcommand{\argmin}{\operatornamewithlimits{argmin }}
\newcommand{\argmax}{\operatornamewithlimits{argmax }}

\def\BibTeX{{\rm B\kern-.05em{\sc i\kern-.025em b}\kern-.08em
    T\kern-.1667em\lower.7ex\hbox{E}\kern-.125emX}}

\begin{document}

\title{Mobile authentication of copy detection patterns: how critical is to know fakes? \\ 
\thanks{S. Voloshynovskiy is a corresponding author.}
\thanks{This research was partially funded by the Swiss National Science Foundation SNF No. 200021\_182063.}
}

\author{\IEEEauthorblockN{Olga Taran, Joakim Tutt, Taras Holotyak, Roman Chaban, Slavi Bonev and Slava Voloshynovskiy}
\IEEEauthorblockA{Department of Computer Science, University of Geneva, Switzerland \\
\{olga.taran, joakim.tutt, taras.holotyak, roman.chaban, slavi.bonev, svolos\}@unige.ch}
}

\maketitle

\begin{abstract}
Protection of physical objects against counterfeiting is an important task for the modern economies. In recent years, the high-quality counterfeits appear to be closer to originals thanks to the rapid advancement of digital technologies. To combat these counterfeits, an anti-counterfeiting technology based on hand-crafted randomness implemented in a form of copy detection patterns (CDP) is proposed enabling a link between the physical and digital worlds and being used in various brand protection applications. The modern mobile phone technologies make the verification process of CDP easier and available to the end customers. Besides a big interest and attractiveness, the CDP authentication based on the mobile phone imaging remains insufficiently studied. In this respect, in this paper we aim at investigating the CDP authentication under the real-life conditions with the codes printed on an industrial printer and enrolled via a modern mobile phone under the regular light conditions. The authentication aspects of the obtained CDP are investigated with respect to the four types of copy fakes. The impact of fakes’ type used for training of authentication classifier is studied in two scenarios: 
\textit{(i)} supervised binary classification under various assumptions about the fakes and \textit{(ii)} one-class classification under unknown fakes. The obtained results show that the modern machine-learning approaches and the technical capacity of modern mobile phones allow to make the CDP authentication under unknown fakes feasible with respect to the considered types of fakes and code design. 
 \end{abstract}


\section{Introduction}

Nowadays, counterfeiting and piracy are among the main challanges for modern economy.
Counterfeiting of medical supplies, food, cosmetics, mechanical parts and goods in general poses tremendous risks to public welfare and health, businesses and brand value reputation. At the same time, many traditional anti-counterfeiting technologies become quickly obsolete in view of rapid technological progress that offers a wide range of modern high-tech tools to the counterfeiters such as modern machine learning systems, high quality digital industrial printers and scanners. On the other hand, many new approaches to anti-counterfeiting appear thanks to the advancement of modern mobile technologies and machine learning algorithms.

\begin{figure}[t!]
	\centering
    \includegraphics[width=0.95\linewidth]{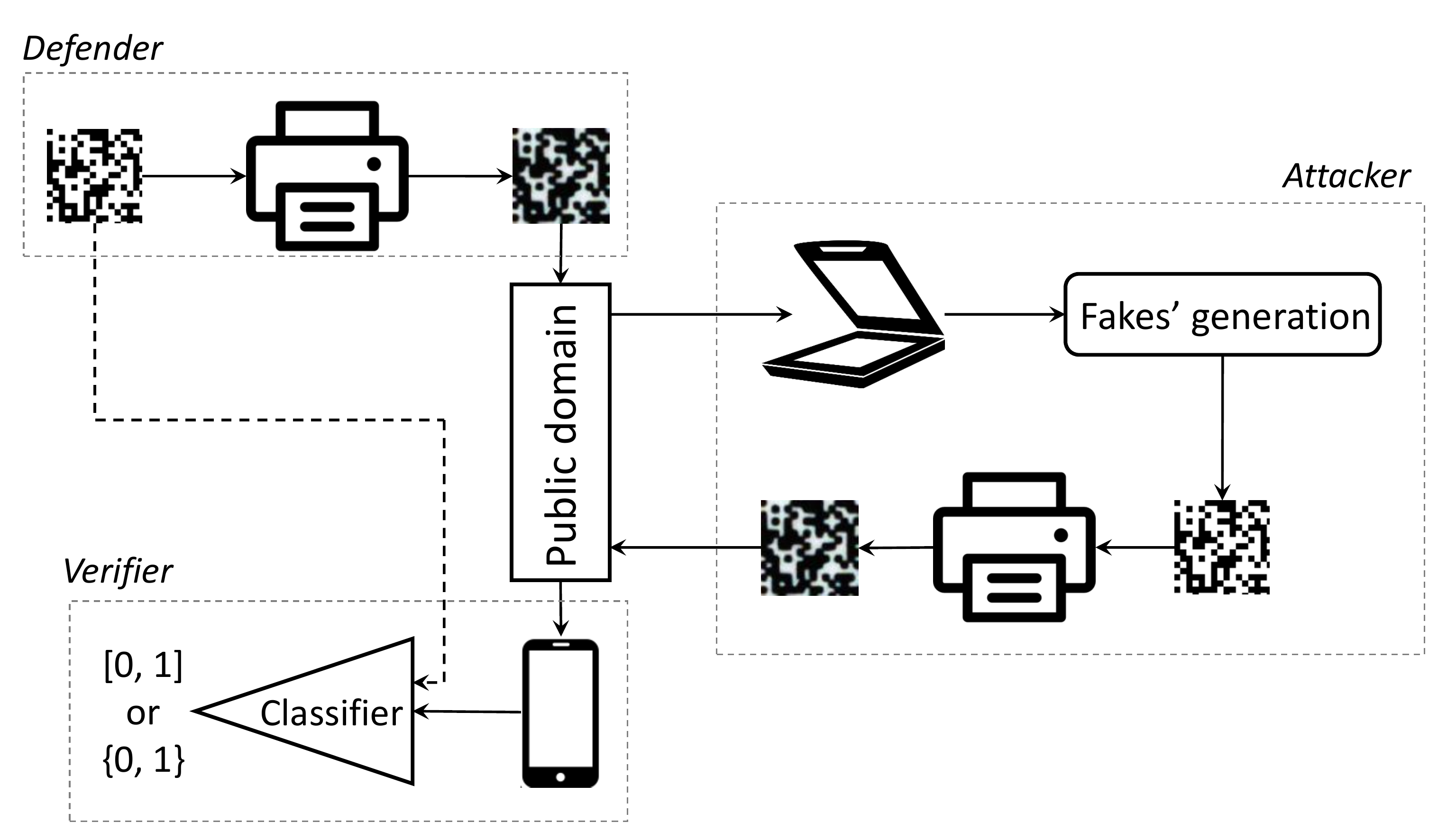}
    \caption{General scheme of the CDP life cycle: \textit{(i)} the generated digital templates are printed by a defender and go to the public domain; \textit{(ii)} an attacker having an access to the publicly available codes can produce different types of fakes that are then also distributed in the public domain; \textit{(iii)} a verifier digitizes the printed codes from the public domain and validates them via a parameterized classifier that might produce either a hard decision (fake/authentic $\sim$ 0/1) or a kind of a sort decision ranging from 0 to 1. The validation might be produced with or without taking the digital templates into account. For the defender-verifier pair the protection problem consists in the minimization of probability of error as a function of the CDP design, used printing and acquisition technologies and used classifier. For the attacker the goal is to maximize the probability of error as a function of the attack construction.}
    \label{fig:pgc life cycle}
\end{figure}

In the recent years, the \textit{printable graphical codes} (PGC) attracted a lot of attention as a link between the physical and digital worlds, which is of great interest for the internet of things, track and trace and brand protection applications. The anti-counterfeiting technology based on the PGC belongs to a family of hand-crafted physical unclonable functions (PUFs) \cite{voloshynovskiy2016physical}. Quite often, the PGC represent a union of the 2D bar codes that are clonable but have a semantic meaning and so-named \textit{copy detection patterns}\cite{picard2004digital} (CDP) that are sensitive to the illegal copying. They might be combined in many different ways \cite{picard2021counterfeit, tkachenko2016printed, nguyen2017watermarking}. The PGC life cycle diagram is schematically shown in Fig. \ref{fig:pgc life cycle}.

\begin{table*}[t!]
	\centering
	\caption{An overview of the existing datasets of CDP: the datasets (1) and (2) are publicly available state-of-the-art datasets and the dataset (3) is created and investigated in the present paper.}
	\label{tab:pgc datasets overview}
	{\small
	\begin{tabular}{l>{\centering\arraybackslash}m{0.15\linewidth}
	ccc>{\centering\arraybackslash}m{0.12\linewidth}} \hline
	\# & Name & Digital templates & Printing & Acquisition & \# of codes \\ \hline
    %
    %
	 (1) & CSGC \cite{yadav2019estimation} & \multicolumn{1}{m{0.15\linewidth}}{ 
            \begin{itemize}[leftmargin=0cm,noitemsep,topsep=0pt]
                \item[] size: $100 \times 100$
                \item[] symbol size: $1 \times 1$
            \end{itemize}  
            }    
		& \multicolumn{1}{m{0.2\linewidth}}{
			\textit{Laser}, at 600 dpi: 
        	\begin{itemize}[leftmargin=0.2cm,noitemsep,topsep=0pt]
        		\item[] $\boldsymbol{\cdot}$ Xerox Phaser 6500
    	    \end{itemize}
            }    
		& \multicolumn{1}{>{\centering\arraybackslash}m{0.18\linewidth}}{ 
            \begin{itemize}[leftmargin=0.2cm,noitemsep,topsep=1pt]
            	\item[] \textit{Scanner}:            
                \item[] $\boldsymbol{\cdot}$ Epson V850 Pro
                \item[] $\;\;$ at 2400 ppi
                \item[] $\;\;$ at 4800 ppi
                \item[] $\;\;$ at 9600 ppi                                
            \end{itemize}  
            }  
		& \multicolumn{1}{m{0.1\linewidth}}{ 
            \begin{itemize}[leftmargin=0cm,noitemsep,topsep=0pt]
                \item[] digital: 950
                \item[] original: 2850
                \item[] fakes: 0
                \item[] total: 3800
            \end{itemize}  
            }                    
     \\[-0.35cm] \hline
     %
     %
	(2) & DP1C \& DP1E \cite{Taran2020icassp} & \multicolumn{1}{m{0.15\linewidth}}{ 
            \begin{itemize}[leftmargin=0cm,noitemsep,topsep=0pt]
                \item[] size: $384 \times 384$
                \item[] symbol size: $6 \times 6$                
            \end{itemize}  
            } 
		& \multicolumn{1}{m{0.2\linewidth}}{
			\textit{Laser}, at 1200 dpi: 
        	\begin{itemize}[leftmargin=0.2cm,noitemsep,topsep=0pt]
        		\item[] $\boldsymbol{\cdot}$ Samsung Xpress 430
	        	\item[] $\boldsymbol{\cdot}$ Lexmark CS310
    	    \end{itemize}
			\textit{Inkjet}, at 1200 dpi: 
        	\begin{itemize}[leftmargin=0.1cm,noitemsep,topsep=0pt]
                \item[] $\boldsymbol{\cdot}$ Canon PIXMA  iP7200
                \item[] $\boldsymbol{\cdot}$ HP OfficeJet Pro 8210
    	    \end{itemize}	    	    			
            } 
		& \multicolumn{1}{>{\centering\arraybackslash}m{0.18\linewidth}}{ 
            \begin{itemize}[leftmargin=0.2cm,noitemsep,topsep=0pt]
            	\item[] \textit{Scanner}:            
                \item[] $\boldsymbol{\cdot}$ Canon 9000F 
                \item[] $\;\;$ at 1200 ppi
                \item[] $\boldsymbol{\cdot}$ Epson V850 Pro
                \item[] $\;\;$ at 1200 ppi
            \end{itemize}  
            }  
		& \multicolumn{1}{m{0.1\linewidth}}{ 
            \begin{itemize}[leftmargin=0cm,noitemsep,topsep=0pt]
                \item[] digital: 384
                \item[] original: 3072
                \item[] fakes: 3072
                \item[] total: 6528
            \end{itemize}  
            }  
     \\[-0.35cm] \hline  
     %
     %
 	 (3) & Indigo mobile  & \multicolumn{1}{m{0.15\linewidth}}{ 
            \begin{itemize}[leftmargin=0cm,noitemsep,topsep=0pt]
                \item[] size: $330 \times 330$
                \item[] symbol size: $5 \times 5$                
            \end{itemize}  
            }  
		& \multicolumn{1}{m{0.2\linewidth}}{
			\textit{Industrial}, at 812 dpi: 
        	\begin{itemize}[leftmargin=0.2cm,noitemsep,topsep=0pt]
        		\item[] $\boldsymbol{\cdot}$ HP Indigo 5500 DS
    	    \end{itemize}
            } 
		& \multicolumn{1}{>{\centering\arraybackslash}m{0.18\linewidth}}{ 
            \begin{itemize}[leftmargin=0.2cm,noitemsep,topsep=0pt]
            	\item[] \textit{Mobile phone}:            
                \item[] $\boldsymbol{\cdot}$ iPhone XS 
                \item[] $\;\;$ auto settings
            \end{itemize}  
            }  
		& \multicolumn{1}{m{0.1\linewidth}}{ 
            \begin{itemize}[leftmargin=0cm,noitemsep,topsep=3pt]
                \item[] digital: 300
                \item[] original: 300
                \item[] fakes: 1200
                \item[] total: 1800
            \end{itemize}  
            }  
     \\[-0.35cm] \hline
                   
	\end{tabular}
	}
\end{table*}

Up to our best knowledge, the CDP based authentication of the PGC under the conditions close to the real-life, where the codes are printed on an industrial printer and enrolled via the modern mobile phones has not been investigated in the prior art publications.  In this respect, in this paper we perform this analysis for several types of copy attacks. Moreover, in view of the fact that a particular type of fakes is unknown to the defender at the test stage, we propose and study an authentication system trained without knowledge of fakes and compare its performance with one that is trained with the complete knowledge of fakes.

Taking into account ethical and non-competition aspects of the considered problem with respect to several competitive technologies on the market, the investigation is performed on the CDP generated based on an open international standard ISO/IEC 16022 \cite{ISO2006}. The main goal is to demonstrate a general approach applicable to the majority of CDP designed with the identical modulation principles rather than to investigate the authentication aspects of some particular CDP.

The main contributions of this paper are:
\begin{itemize}
    \item A new dataset of CDP, which is produced on the industrial printing equipment HP Indigo 5500 DS and is acquired on the mobile phone iPhone XS under the regular light conditions.
    \item Investigation of the authentication aspects of CDP with respect to the typical copy fakes in a supervised and one-class classification setups.
    \item Analysis of the supervised and one-class classification of the CDP from the information theory point of view. 
\end{itemize}

	



\noindent\paragraph*{Notations} We use the following notations: $\t \in \{0, 1\}^{m \times m}$ denotes an original digital template representing CDP; $\x \in \mathbb{R}^{m \times m}$ corresponds to the image of the original printed code, while $\f \in \mathbb{R}^{m \times m}$ is used to denote the image taken from a printed fake code; $\y \in \mathbb{R}^{m \times m}$ stands for a probe that might be either original or fake. $p_t(\t)$ and $p_\D(\x)$ correspond to the data distributions of the digital templates and original printed codes, respectively. A discriminator corresponding to the Kullback–Leibler divergence is denoted as $\Dx$, where the subscript indicates the space to which this discriminator is applied to. 


\section{Indigo mobile dataset}
\label{sec: dataset}

Despite a big recent popularity of CDP, there are not so many public datasets available for investigation and reproducible research. The CDP dataset production is a time consuming and very costly process. Thus, the majority of the research experiments are performed either on synthetic data or on small private datasets. This also partially explains the lack of complete understanding about the clonability and performance of CDP under different classes of attacks.

\begin{figure}[t!]
	\centering
	\begin{subfigure}{0.15\textwidth}     
		\centering	
        \includegraphics[width=1\linewidth,valign=t]{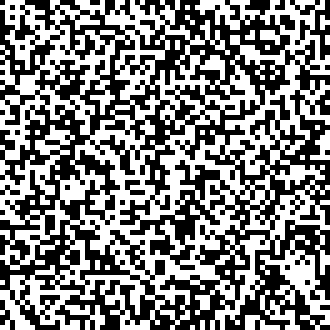}
        \caption{Digital template.}
        \label{fig:digital template}
	\end{subfigure}
	\begin{subfigure}{0.15\textwidth}
		\centering
        \includegraphics[width=1\linewidth,valign=t]{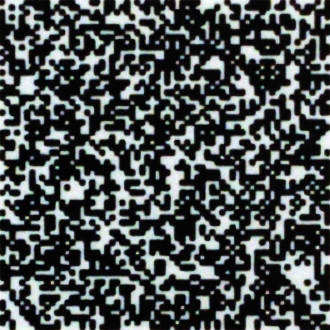}
        \caption{Original.}
        \label{fig:original}
	\end{subfigure}
	\begin{subfigure}{0.15\textwidth}
		\centering
        \includegraphics[width=1\linewidth,valign=t]{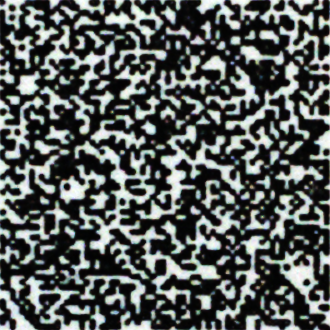}
        \caption{Fakes \#1 white.}
        \label{fig:f1w}
	\end{subfigure}	
	\\
	\begin{subfigure}{0.15\textwidth}
		\centering
        \includegraphics[width=1\linewidth,valign=t]{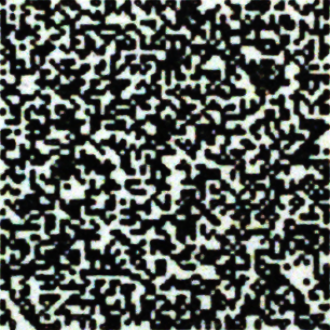}
        \caption{Fakes \#1 gray.}
        \label{fig:f1g}
	\end{subfigure}	
	\begin{subfigure}{0.15\textwidth}
		\centering
        \includegraphics[width=1\linewidth,valign=t]{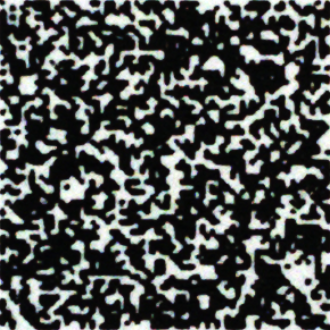}
        \caption{Fakes \#2 white.}
        \label{fig:f2w}	
    \end{subfigure}    
	\begin{subfigure}{0.15\textwidth}
		\centering
        \includegraphics[width=1\linewidth,valign=t]{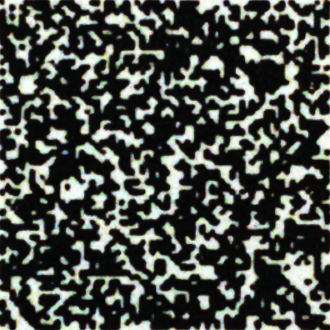}
        \caption{Fakes \#2 gray.}
        \label{fig:f2g}
	\end{subfigure}	
    \caption{Examples of original and fake codes taken by a mobile phone from the Indigo mobile dataset. }
    \label{fig:auth_db codes visualisation}        
\end{figure}

Up to our best knowledge, there are only few public CDP datasets: \textit{(i)} \textit{DP0E} \cite{Taran2019icassp} and its extensions \textit{DP1C} \& \textit{DP1E} \cite{Taran2020icassp} and \textit{(ii)} \textit{CSGC} \cite{yadav2019estimation}. The datesets' details are summarized in Table \ref{tab:pgc datasets overview}. These state-of-the-art datasets were created to investigate the clonability aspects of CDP. Thus, the codes were printed on the desktop printers and enrolled by the scanners at the high resolution. At the same time, these conditions are not suitable to investigate the authentication of the CDP in the industrial settings, which is of great practical importance. In this respect, we present a new dataset, named \textit{Indigo mobile} dataset that contains 300 distinct digital DataMatrix \cite{ISO2006} templates $\t \in \{0, 1\}^{330 \times 330}$ with the symbols size of $5 \times 5$\footnote{To ensure accurate symbol representation, each printed symbol should be represented by at least $3 \times 3$ pixels. The anti-counterfeiting is an important problem for the developing countries, where the relatively cheep phones with a low resolution (about 600 - 900 ppi) predominate. Taking into account the difference in the printing (812 dpi) and potential acquisition resolution (600 - 900 ppi)  the symbols size should be about 4x4 or even 5x5.}. To simulate the real life scenario, the codes are printed on the industrial printer HP Indigo 5500 DS at the resolution 812 dpi.
The acquisition of the printed codes is performed under regular room light using mobile phone iPhone XS under the automatic photo shooting settings at the resolution 12 Megapixels. The photos are taken in DNG format to avoid built-in mobile phone image post-processing. The final codes are converted to the RGB format based on the publicly available code \cite{cv2.cvtColor}\footnote{Despite the visually black and white nature of the CDP the authentication based on codes taken by the mobile phone in color mode is more efficient compared to the grayscale mode due to the different sensitivity of the color channels and corresponding degradation caused by converting a three-channels color image into a single-channel grayscale one.}. Examples of digital template and corresponding enrolled printed original code are given in Fig. \ref{fig:digital template} and \ref{fig:original}, respectively. All codes are synchronized using additional synhromarkers ensuring the correspondence between the digital templates and acquired codes. 

As the most typical scenario for an unexperienced counterfeiter simulation we produce copy fakes based on the two standard copy machines (in a copy regime "text"): \textit{(1)} RICOH MP C307 and \textit{(2)} Samsung CLX-6220FX. The copy attacks based on advanced machine learning are addressed  in \cite{Taran2020icassp, yadav2019estimation} and are not considered in this paper. The fakes are produced on the white and gray paper (80 g/m\textsuperscript{2}). Thus, for each original printed code we produce four fake codes: 
\begin{enumerate}[noitemsep] 
	\item \textit{Fakes \#1 white}: made by the copy machine \textit{(1)} on the white paper.
	\item \textit{Fakes \#1 gray}: made by the copy machine \textit{(1)} on the gray paper.
	\item \textit{Fakes \#2 white}: made by the copy machine \textit{(2)} on the white paper.
	\item \textit{Fakes \#2 gray}: made by the copy machine \textit{(2)} on the gray paper.
\end{enumerate}

\begin{table*}[t!]
	\centering
	\renewcommand*{\arraystretch}{1.25}
	\caption{The average (over five runs) classification error in \% of the supervised binary classifier.}
	\label{tab:supervised classification}
	{\small
	\begin{tabular}{lccccc} \hline
	%
	%
	Setup  & Original ($P_{miss}$) & Fake \#1 white ($P_{fa}$) & Fake \#1 gray ($P_{fa}$) & Fake \#2 white ($P_{fa}$) & Fake \# 2 gray ($P_{fa}$) \\	\hline
	(1) All fakes & 0 & 0 & 0 & 0 & 0\\
	(2) Fakes \#1 white & 0 & 0 & 0.14 ($\pm$0.32) & 0 & 0 \\
	(3) Fakes \#1 gray  & 0 & 0 & 0 & 0 & 0 \\
	(4) Fakes \#2 white & 0 & 99.43 ($\pm$0.32) & 100 & 0 & 0 \\
	(5) Fakes \# 2 gray & 0 & 99.29 ($\pm$0.5) & 99.86 ($\pm$0.32) & 0 & 0 \\ \hline
	\end{tabular}
	}
\end{table*}

The acquisition of the produced fakes is done in the same way on the same mobile phone under the same photo and light settings as for the original codes. Examples of the produced fakes are given in Fig. \ref{fig:f1w} - \ref{fig:f2g}. It is important to note that the fakes \#1 visually closely resemble the original codes, while the fakes \#2 have bigger dot gain and visually are more different from the original codes.

As it is summarized in Table \ref{tab:pgc datasets overview}, the Indigo mobile dataset contains 1800 images of codes: \textit{(i)} 300 distinct digital templates; \textit{(ii)} 300 enrolled original printed codes and \textit{(iii)} 1200 enrolled fake printed codes (300 originals × 4 type of fakes)\footnote{The Indigo mobile dataset is available at \url{https://github.com/sip-group/snf-it-dis/tree/master/datasets/indigomobile}.}.

For the simulation purposes, the Indigo mobile dataset was split into three subsets: \textit{(1)} the training with 40\% of data, \textit{(2)} the validation with 10\% of data and \textit{(3)} 50\% of data is used for the test. Taking into account a relatively small amount of data available for the training the next data augmentations are used: \textit{(i)} the rotations on 90\degree, 180\degree and 270\degree; \textit{(ii)} the gamma correction with variable function $(.)^\gamma$, where $\gamma$ is the parameter of gamma correction.

\section{Supervised classification}
\label{sec:supervised classification}

\subsection{Theoretical analysis}

In this paper we consider the supervised classification as a base-line scenario to validate the mobile phone based authentication of the CDP. The complete availability of fakes at the training stage in the supervised classification gives the defender an information advantage over the attacker.

To link the supervised classification problem with the considered one-class classification operating under the absence of fakes, we introduce a common theoretical basis based on an information-theoretic formulation. The problem of a supervised classifier training given the labeled data $\{\x_i, \c_i\}^N_{i=1}$ generated from a joint distribution $p(\x, \c)$ is formulated as a training of a parameterized network $p_\bphi(\c|\x)$ that is an approximation of $p(\c|\x)$ originating from the chain rule decomposition $p(\x, \c) = p_\D(\x) p(\c|\x)$. The training of the network $p_\bphi(\c|\x)$ is performed based on the maximisation of a mutual information $I_{\bphi}(\X;\C)$ between $\x$ and $\c$ via $p_\bphi(\c|\x)$:

\begin{equation} 
    \hat{\bphi} = \argmax_{\bphi} I_{\bphi}(\X;\C),
    \label{eq:ib supervised classification max}     
\end{equation}
%
that can be rewritten as: 
%
\begin{equation} 
    \hat{\bphi} = \argmin_{\bphi}  \loss_{\textrm{Supervised}}(\bphi ),
    \label{eq:ib supervised classification min}     
\end{equation}
%
where $\loss_{\textrm{Supervised}}(\bphi ) =  - I_{\bphi}(\X;\C)$.


The mutual information in (\ref{eq:ib supervised classification max}) is find as: 
%
\begin{equation}
	\begin{aligned} 
		I_\bphi(\X; \C) & \triangleq 
		%
		\mathbb{E}_{p(\x,\c)} \left[ \log \frac{p_\bphi(\c|\x)}{p_c(\c)}  \right] \\
		%
		& =  \underbrace{\mathbb{E}_{p(\x,\c)} \left[ \log p_\bphi(\c|\x) \right] }_\text{$\Dcc$} - \underbrace{\mathbb{E}_{p_c(\c)} \left[ \log p_c(c)  \right]}_\text{= constant},
	 \end{aligned} 
	\label{eq:ib-occ first term direct decomposition}
\end{equation}
%
where $H(\C) = -\mathbb{E}_{p_c(\c)} \left[ \log p_c(c)  \right]$ is an entropy of $\c$ and it is a constant that does not depend on $\bphi$. 

The optimisation problem (\ref{eq:ib supervised classification min}) reduces to: 

\begin{equation} 
\begin{aligned}
\hat{\bphi} & = \argmin_{\bphi}  \loss_{\textrm{Supervised}}(\bphi) = \argmin_{\bphi} -\Dcc.
\end{aligned}
\label{eq:IB_sup_class_final_optimization}
\end{equation}


\subsection{Empirical results}

\begin{figure*}[t!]
	\centering

	\begin{subfigure}{0.19\textwidth}     
		\centering	
    \includegraphics[width=1.13\linewidth]{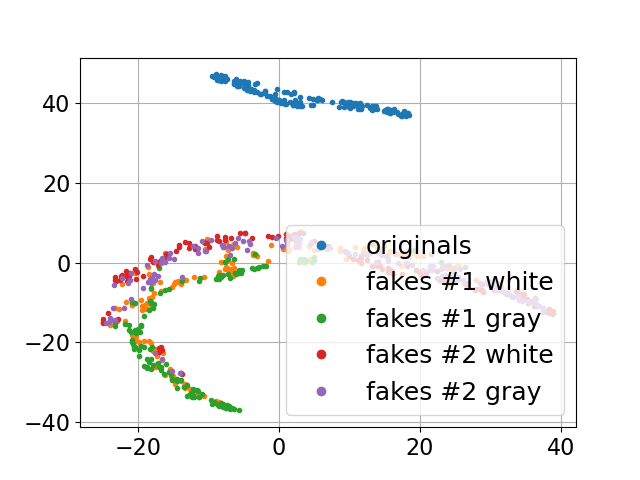}
        \caption{}
        \label{fig:tsne all fakes}
	\end{subfigure}
	\begin{subfigure}{0.19\textwidth}     
		\centering	
    \includegraphics[width=1.13\linewidth]{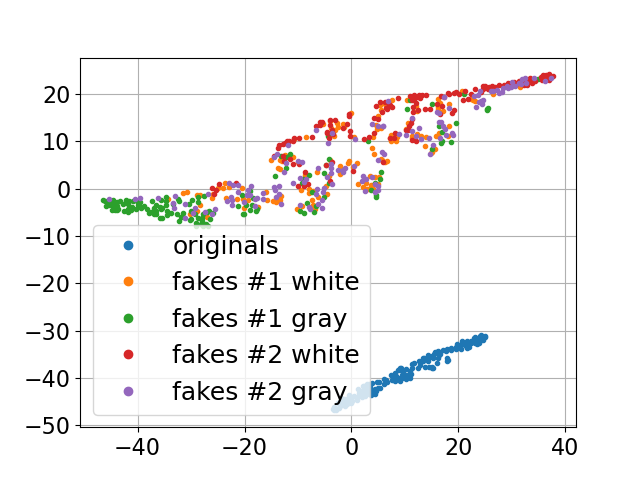}
        \caption{}
        \label{fig:tsne f1w}
	\end{subfigure}
	\begin{subfigure}{0.19\textwidth}
		\centering
    \includegraphics[width=1.13\linewidth]{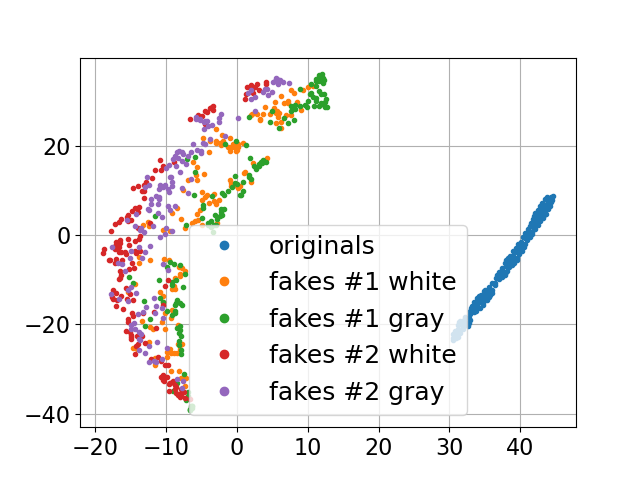}
        \caption{}
        \label{fig:tsne f1g}
	\end{subfigure}
	\begin{subfigure}{0.19\textwidth}     
		\centering	
    \includegraphics[width=1.13\linewidth]{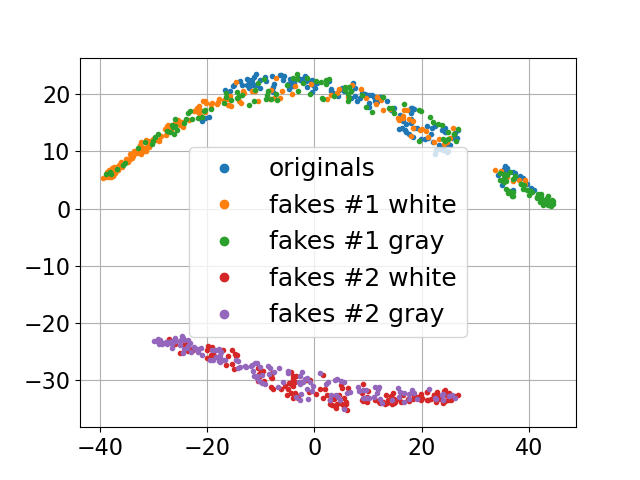}
        \caption{}
        \label{fig:tsne f2w}
	\end{subfigure}
	\begin{subfigure}{0.19\textwidth}
		\centering
    \includegraphics[width=1.13\linewidth]{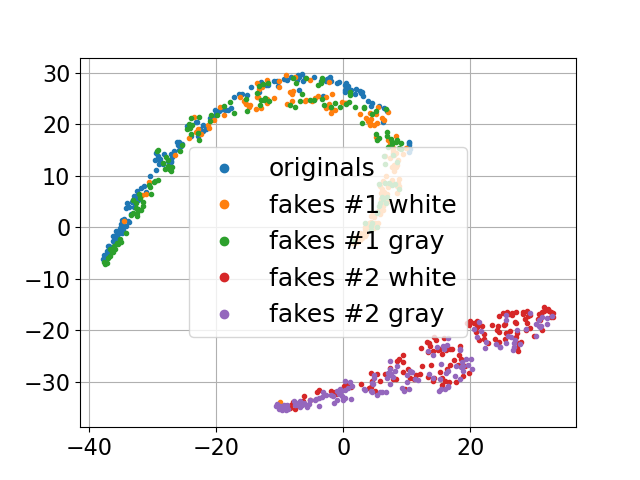}
        \caption{}
        \label{fig:tsne f2g}
	\end{subfigure}

    \caption{The t-SNE visualization of the last layer before an activation function of the supervised binary classifier trained on the originals and (a) all type of fakes, (b) fakes \#1 white, (c) fakes \#1 gray, (d) fakes \#2 white, (e) fakes \#2 gray.}
    \label{fig:tsne binary classification}        
\end{figure*}		


The main goal of this section is to investigate the influence of the different types of fakes on the authentication accuracy under the supervised classification. In this respect, the binary classifier was trained in five different setups\footnote{To avoid the bias in the choice of training and test data, in each setup the classifier was trained five times on the randomly permuted data. Each time the learning rate equaled to 1e-4, the batch size was 21 and the cross-entropy loss was used. The Adam was used as an optimizer. The gamma correction was performed with $\gamma \in [0.4, 1.3]$ with step $0.2$. The more details about the used data augmentations are given in Section \ref{sec: dataset}.}  on the original codes and (1) all type of fakes, (2) fakes \#1 white, (3) fakes \#1 gray, (4) fakes \#2 white, (5) fakes \#2 gray.


At the inference stage, the query sample $\y$, which might be either original $\x$ or one of the fakes $\f^k$, $k = 1, ..., 4$, is passed through the deterministic classifier $g_\bphi$ such that $p_\bphi(\c|\x) = \delta(\c - g_\bphi(\x))$ and $\delta(.)$ denotes the Dirac delta-function or simply $\c = g_\bphi(\x)$. Herewith, $g_\bphi$ is trained with respect to the term $\Dcc$ in (\ref{eq:IB_sup_class_final_optimization}).
The classification accuracy is evaluated  with respect to the probability of miss $P_{miss}$ and the probability of false acceptance $P_{fa}$: 
\begin{equation}
\left \{
\begin{array}{lll}
P_{fa}   & = & \textrm{Pr}\{ g_{\bphi}(\Y) = \c_1   \;|\; \mathcal{H}_0 \}, \\
P_{miss} & = & \textrm{Pr}\{ g_{\bphi}(\Y) \neq \c_1 \;|\; \mathcal{H}_1 \}, 

\end{array}
\right.
\label{eq:4}
\end{equation}
where $\c_1$ denotes a class of original codes,  $\mathcal{H}_1$ corresponds to the hypothesis that the query $\y$ is an original code and $\mathcal{H}_0$ is the hypothesis that the query $\y$ is a fake code.

The obtained classification error is given in Table \ref{tab:supervised classification}. When all types of the fakes are available for training that corresponds to the setup (1) the classifier distinguishes the original codes with high accuracy even with respect to the most proximate fakes \#1. This can also be seen in Fig. \ref{fig:tsne all fakes}, where the t-SNE \cite{hinton2002stochastic} visualisations of the last layer before an activation function of corresponding trained classifier is shown. From this visualisation it is possible to observe two main clusters: the first one is formed by the original codes (blue cluster) and the second one is formed by the fakes (multi-color cluster). In the setups (2) and (3), where the classifier is trained only on the original codes and the fakes \#1 (white - the setup (2) or gray - the setup (3)) the authentication error is low even with respect to the unseen fakes \#2. The latent space visualisations presented in Fig. \ref{fig:tsne f1w} and \ref{fig:tsne f1g} are similar to the setup (1). In the setups (4) and (5), the situation is different: if during the training the classifier does not observe the high quality fakes (which correspond to the fakes \#1 in our dataset)  and is trained only on the fakes that are far away from the manifold of original data (in our dataset they correspond to fakes \# 2), then the authentication of unseen high quality fakes becomes more complicated and less accurate as shown in Table \ref{tab:supervised classification}. The latent space visualisations (Fig. \ref{fig:tsne f2w} and \ref{fig:tsne f2g}) show that the fakes \#1 form one cluster with the original codes and are almost certainly authenticated as the genuine codes.


\section{One-class classification}

\begin{figure}[t!]
    \centering
	\includegraphics[width=0.75\linewidth]{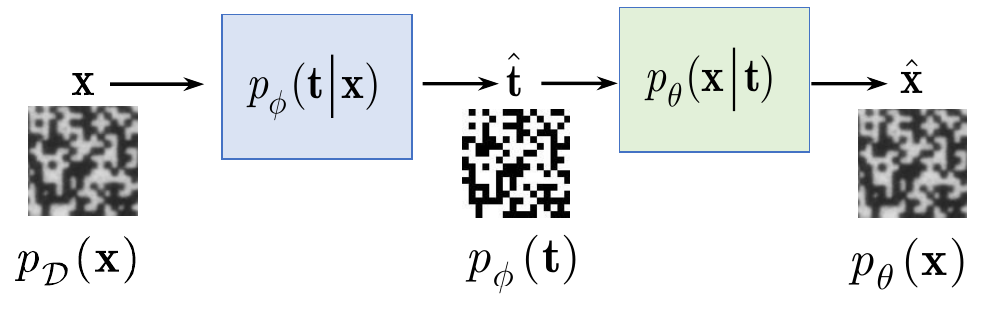}
	\caption{General scheme of a deep model that aims at estimating the digital templates $\hat{\t}$ from the original printed codes $\x$ with the following mapping of the estimated digital templates $\hat{\t}$ back to the printed codes $\hat{\x}$.}
	\label{fig:ib one class classification}	
\end{figure}

\subsection{Theoretical analysis}

The supervised classification scenario is an ideal case for the informed defender but it requires the knowledge of the corresponding fakes. With respect to the CDP authentication, the fakes' collection might be quite expensive and time consuming process. Moreover, taking into account the permanent improvement of technologies available for the fakes' production, it is very difficult to predict in advance what kind of fakes will be used by the attackers. In this respect, the one-class classification problem, where the authentication model is trained only on the original data disregarding the potential fakes, is of great practical importance.

In general case, it is possible to highlight the two main parts of the considered one-class classification system: \textit{(1)} feature extraction and \textit{(2)} one-class classifier. The one-class classifier by itself is an important part of the whole process. However, the main focus of this work is to find a set of reliable features that allow to distinguish between the original and fake codes even by using simple one-class classifiers. In this respect, we use the one-class support vector machine (OC-SVM) \cite{chen2001one} as a one-class classifier model. Alternatively, one can use an one-class deep classifier \cite{ruff2018deep}.

\begin{table*}[t!]
	\centering
	\renewcommand*{\arraystretch}{1.25}
	\caption{The average (over five runs) authentication error in \% of the one-class classification based on the OC-SVM.}
	\label{tab:ib oc classification results}
	{\small
	\begin{tabular}{cccccccc} \hline
		\multirow{2}{*}{Setup} & Feature & OC-SVM & Original & Fake \#1 white & Fake \#1 gray & Fake \#2 white & Fake \#2 gray \\ 
		& extractor & input & $P_{miss}$ & $P_{fa}$ & $P_{fa}$ & $P_{fa}$ & $P_{fa}$ \\[0.1cm] \hline 
		
		(1) & $\mathcal{L}_\textrm{One-class}^1$ & $\{ \Dtt, \Dxx \}$ & 0.28 ($\pm$0.39) & 0 & 0 & 0 & 0 \\
		
		(2) & $\mathcal{L}_\textrm{One-class}^2$ & $\{ \Dtt, \Dt \}$ & 40.57 ($\pm$54.26) & 0.57 ($\pm$0.59) & 0.57 ($\pm$0.92) & 0 & 0  \\ 
		
		(3) & $\mathcal{L}_\textrm{One-class}^2$ & $\{ \Dxx, \Dx \}$ & 4.26 ($\pm$3.09) & 0 & 0 & 2.55 ($\pm$3.08) & 3.26 ($\pm$4.85) \\
		
		(4) & $\mathcal{L}_\textrm{One-class}^2$ & $\{ \Dtt, \Dxx \}$ & \textbf{0.14 ($\pm$0.32)} & \textbf{0} & \textbf{0} & \textbf{0} & \textbf{0} \\ 
		
		(5) & $\mathcal{L}_\textrm{One-class}^2$ & $\{ \Dtt, \Dt, \Dxx, \Dx \}$ & 3.55 ($\pm$1.66) & 0 & 0 & 0 & 0 \\ \hline		
	\end{tabular}
	}
\end{table*}    

As a feature extractor we investigate a deep auto-encoding model $\x \to \hat{\t} \to \hat{\x}$, where $\hat{\t}$ is considered as a latent space representation as shown in Fig. \ref{fig:ib one class classification}.
The loss-function for the considered feature extracting system is: 
%
\begin{equation} 
	\loss_{\textrm{One-class}}(\bphi, \btheta) =  - I_\bphi(\X; \T) - \beta I_{\bphi,\btheta}(\T; \X),
	\label{eq:ib-oc wrt t}
\end{equation}
where $\beta$ controls the relative importance of the two objectives.

\begin{figure}[t!]
	\centering
    \includegraphics[width=1.01\linewidth]{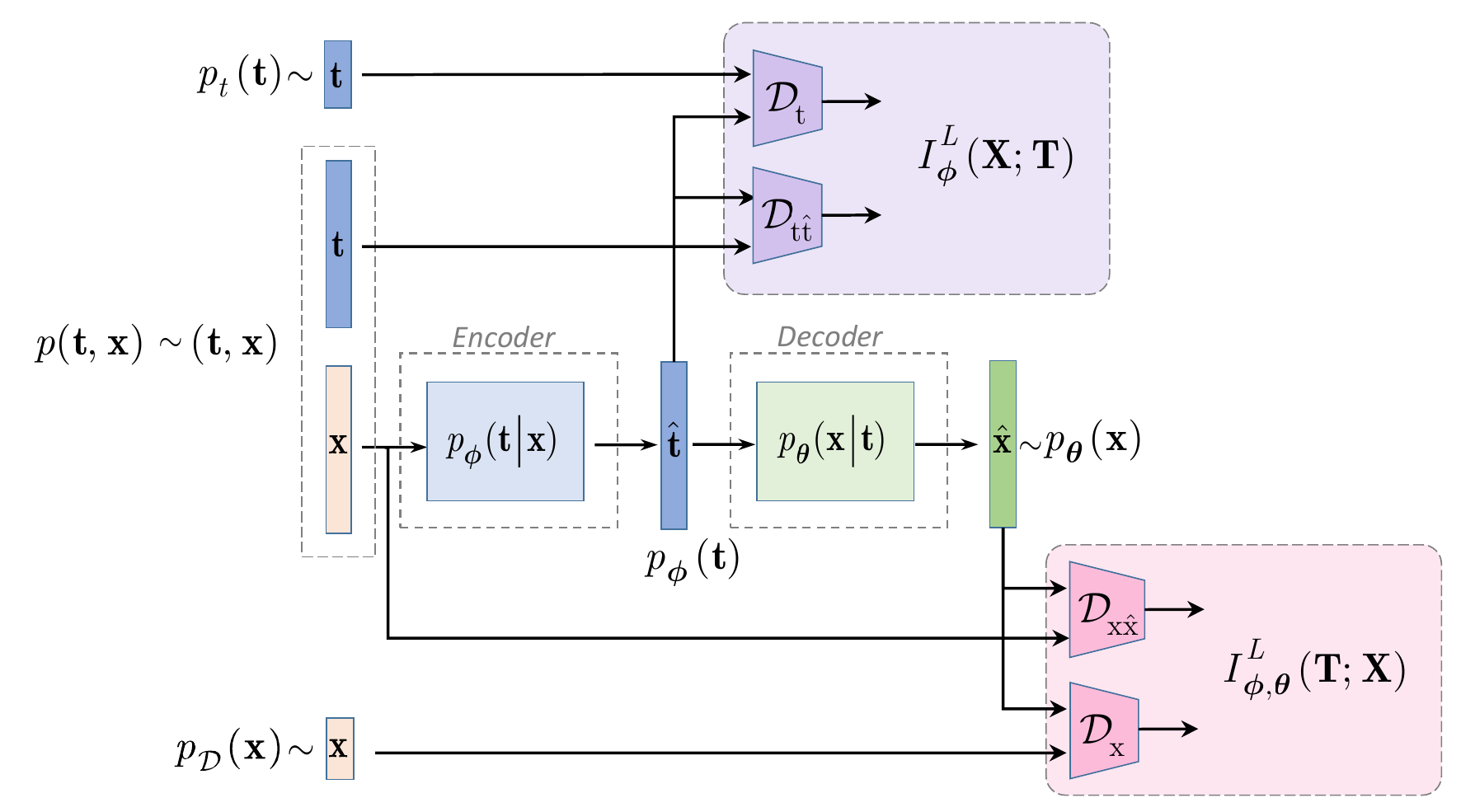}
    \caption{The feature extraction based on the estimation of the digital templates $\hat{\t}$  via $\Dtt$ and $\Dt$ and the printed codes $\hat{\x}$ via $\Dxx$ and $\Dx$ terms.}
    \label{fig:dtt dt dxx dx hc scheme}
\end{figure}

The first mutual information term  in (\ref{eq:ib-oc wrt t}) is defined as $I_\bphi(\X; \T) = \mathbb{E}_{p_\D(\x)} \left[\mathbb{E}_{p_\bphi(\t|\x)} \left[ \log \frac{p_\bphi(\t|\x)}{p_t(\t)}  \right] \right]$. According to \cite{Svolos2020entropy}, using the variational decomposition  it can be lower bounded as $I_\bphi(\X; \T) \geq I^L_{\bphi}(\X;\T)$, where:

\begin{equation}
\begin{aligned} 
I^L_{\bphi}(\X;\T) \triangleq &   
   \underbrace{\mathbb{E}_{p_\D(\x)}  \left[ \mathbb{E}_{p_\bphi(\t|\x)} \left[  \log p_{\bphi}(\t|\x) \right]\right]}_\text{$\Dtt$} \\
   & -
 \underbrace{D_{\mathrm{KL}}\left(p_t(\t) \| p_\bphi(\t)\right)}_\text{$\Dt$},
\end{aligned} 
\label{MI_decoder1}
\end{equation}
%
where $D_{\mathrm{KL}}\left(p_t(\t) \| p_{\bphi}(\t)\right) = \mathbb{E}_{p_t(\t)} \left[ \log \frac{p_t(\t)}{p_{\bphi}(\t)} \right]$.

The second mutual information term in (\ref{eq:ib-oc wrt t}) determined as $I_{\bphi,\btheta}(\T;\X) = \mathbb{E}_{p_\D(\x)} \left[\mathbb{E}_{p_\bphi(\t|\x)} \left[  \log \frac{p_\btheta(\x|\t)}{p_\D(\x)}  \right] \right]$ can be decomposed and bounded in a way similar to the first term: $I_{\bphi,\btheta}(\T;\X) \geq I^L_{\bphi, \btheta}(\T;\X)$, where:
%

\begin{equation}
\begin{aligned} 
I^L_{\bphi,\btheta}(\T;\X) \triangleq  &  
   \underbrace{\mathbb{E}_{p_\D(\x)}  \left[ \mathbb{E}_{p_{\bphi}(\t|\x)} \left[  \log p_{\btheta}(\x| \t) \right]\right]}_\text{$\Dxx$} \\
   & -
 \underbrace{D_{\mathrm{KL}}\left(p_\D(\x ) \| p_{\btheta}(\x)\right)}_\text{$\Dx$},
\end{aligned} 
\label{ib-occ fourth term lower bound}
\end{equation}
%
where $D_{\mathrm{KL}}\left(p_\D(\x ) \| p_{\btheta}(\x)\right) = \mathbb{E}_{p_\D(\x)} \left[ \log \frac{p_\D(\x)}{p_{\btheta}(\x)} \right]$.

Combining the obtained decompositions the final optimization problem schematically shown in Fig. \ref{fig:dtt dt dxx dx hc scheme} is:
\begin{equation} 
\begin{aligned}
(\hat{\bphi}, \hat{\btheta}) & = \argmin_{\bphi, \btheta}  
\loss_\textrm{One-class}^L(\bphi, \btheta) \\
 & = \argmin_{\bphi, \btheta}  - (\Dtt - \Dt) - \beta (\Dxx - \Dx).
\end{aligned}
\label{eq:IB_final_optimization}
\end{equation}

For empirical evaluation of the theoretically obtained features' extractor we consider two basic scenarios of estimation of the digital templates $\hat{\t}$ and the printed codes $\hat{\x}$ based on:
\begin{itemize}
	\item terms $\Dtt$ and $\Dxx$:
	\begin{equation}
		\mathcal{L}^1_\textrm{One-class}(\bphi, \btheta) = - \Dtt - \beta  \Dxx; 
		\label{ch4_eq:oc classification dtt dxx}		
	\end{equation}
	\item terms $\Dtt$, $\Dt$, $\Dxx$ and $\Dx$:
	\begin{equation}
		\mathcal{L}^2_\textrm{One-class}(\bphi, \btheta) = - \Dtt +  \Dt - \beta \Dxx + \beta \Dx.
		\label{ch4_eq:oc classification dtt dt dxx dx}		
	\end{equation}
\end{itemize}

\subsection{Empirical results}

\begin{figure}[t!]
	\centering
    \includegraphics[width=1\linewidth]{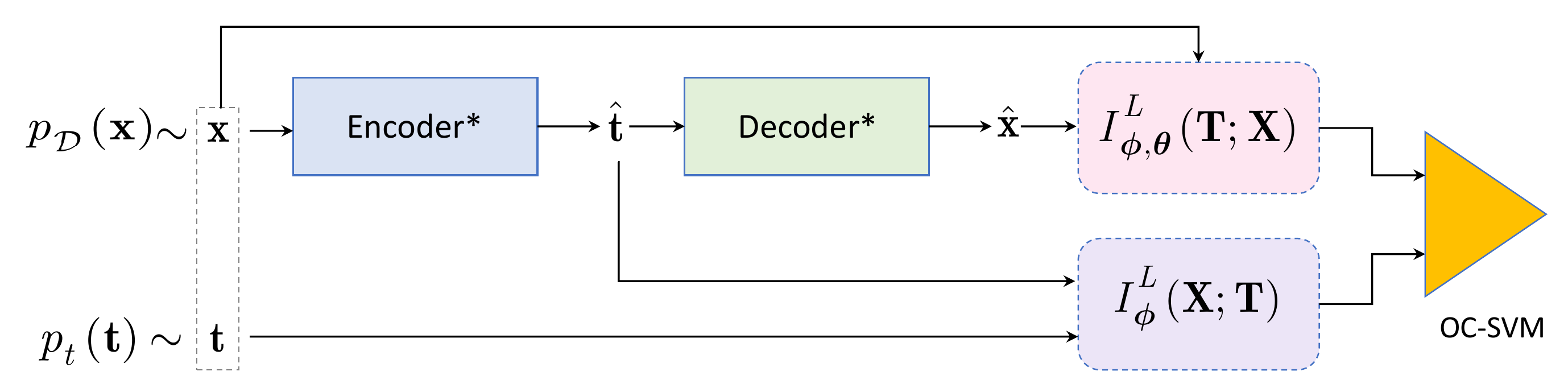}
    \caption{The OC-SVM training procedure.}
    \label{fig:oc classification scheme}
\end{figure}

The general schema of the OC-SVM training is illustrated in Fig. \ref{fig:oc classification scheme}: the encoder and decoder parts of the feature extraction model shown in Fig. \ref{fig:dtt dt dxx dx hc scheme} are pre-trained and fixed (as indicated by a "*")\footnote{The encoder and decoder models were trained with respect to the $\Dtt$ and $\Dxx$ term correspondingly and were based on the U-Net architecture. The KL-divergence terms $\Dt$ and $\Dx$ were implemented in a form of density ratio estimator \cite{goodfellow2014generative}. To avoid the bias in the choice of training and test data, the system was trained five times on the randomly shifted data. Each time the learning rate equaled to 1e-4, the batch of size 18 and the MSE loss were used. The Adam was used as an optimizer. The gamma correction was performed with $\gamma \in [0.5, 1.2]$ with step $0.1$.  The more details about the used data augmentations are given in Section \ref{sec: dataset}.} and the OC-SVM is trained on the different combinations of $\Dtt$ and $\Dt$ terms' outputs that are the results of $I^L_{\bphi}(\X;\T)$ decomposition and the $\Dxx$ and $\Dx$ terms' outputs that are the results of $I^L_{\bphi, \btheta}(\T;\X)$ decomposition\footnote{The OC-SVM was trained to minimize the $P_{miss}$ on the validation sub-set. \\ \indent The python code for both supervised and one-class classification scenarios is available at \url{https://github.com/taranO/Mobile-authentication-of-CDP}.}.

At the inference stage, the query sample $\y$, which might be either original code $\x$ or one of the fakes $\f^k$, $k = 1, ..., 4$, is at first  passed through the feature extractor and then the corresponding feature vector is classified via pre-trained OC-SVM. The classification accuracy is evaluated  with respect to the probability of miss $P_{miss}$ and the probability of false acceptance $P_{fa}$ given in (\ref{eq:4}).

From the obtained authentication results given in Table \ref{tab:ib oc classification results} one can note that the combination of the $\Dtt$ and $\Dxx$ terms' outputs is the most accurate feature vector among the considered ones for the OC-SVM. Moreover, it can be seen that the setup (4)-$\mathcal{L}_\textrm{One-class}^2$ produces the error that is two times smaller than the setup (1)-$\mathcal{L}_1$, although, the same combination of $\Dtt$ and $\Dxx$ terms' outputs are used for the OC-SVM input. It can be explained by the fact that the terms $\Dt$ and $\Dx$ in the $\mathcal{L}_\textrm{One-class}^2$ model play a role of learnable regularizers on the $\Dtt$ and $\Dxx$ terms correspondingly. That allows to make the estimations of the digital templates and printed codes more accurate for the original codes. This in turn leads to the decrease of the authentication error. 

The obtained results demonstrate sufficiently accurate authentication of the CDP with respect to the typical copy attacks by a classifier trained only on the original codes without taking any fakes codes into account.

\section{Conclusions}

In the present work, we investigated the authentication aspects of the modern CDP produced under the conditions close to the real life: the codes are printed on the industrial printer and enrolled via the modern mobile phone under regular light conditions. 

As a base-line we perform a theoretical and empirical evaluation of the supervised binary classification with defined training sets of fakes. The obtained results show  that the fakes used for training play an important role: if classifier observes the high quality fakes at training that are close to manifold of the original codes it is capable to authenticate the unseen middle quality fakes with a high accuracy. In contrast, if during the training the classifier observes only the middle quality fakes it is not capable to authenticate unseen high quality fakes reliably. 

To avoid a problem of search and acquisition of the suitable fakes for the training and theirs quick obsolescence in view of rapid technological progress, we investigated the one-class classification approach that does not require any fakes for the training. The investigated feature extraction system shows promising results even with respect to the one-class classification based on a simple OC-SVM model.

For the future work it is important to investigate the authentication aspects of the modern CDP with respect to the physical references and compare it with the considered authentication with respect to the digital templates. The combination of these two approaches is of great interest too. Finally, it is important to investigate the proposed methods to the CDP authentication with respect to the machine learning fakes. 



\bibliographystyle{IEEEtran}
\bibliography{references.bib}

\end{document}